\documentclass[amsmath,amssymb,superscriptaddress,prl,reprint,widetext,longbibliography]{revtex4-2}

\usepackage[utf8]{inputenc}
\usepackage{graphicx}
\usepackage{xcolor}
\usepackage{hyperref}
\usepackage{notes2bib}
\usepackage{siunitx}
\usepackage{scalerel}
\usepackage{bm}
\usepackage{mathtools}

\newcommand{\probe}{x}
\newcommand{\degr}{\mathbf x}
\newcommand{\prot}{X}
\newcommand{\trap}{X}
\newcommand{\amp}{\hat{X}}

\DeclareSIUnit\Molar{\text{M}}
\DeclareSIUnit\rad{\text{rad}}

\begin{document}

\title{Equilibrium trajectories quantify second order violation of fluctuation dissipation theorem without need of a model}

\author{Juliana Caspers}
\email{j.caspers@theorie.physik.uni-goettingen.de}
\affiliation{Institute for Theoretical Physics, Georg-August-Universit\"{a}t G\"{o}ttingen, 37073 G\"{o}ttingen, Germany}

\author{Karthika Krishna Kumar}
\affiliation{Fachbereich Physik, Universit\"{a}t Konstanz, 78457 Konstanz, Germany}

\author{Clemens Bechinger}
\affiliation{Fachbereich Physik, Universit\"{a}t Konstanz, 78457 Konstanz, Germany}

\author{Matthias Krüger}
\email{matthias.kruger@uni-goettingen.de}
\affiliation{Institute for Theoretical Physics, Georg-August-Universit\"{a}t G\"{o}ttingen, 37073 G\"{o}ttingen, Germany}

\begin{abstract} 
Quantifying and characterizing fluctuations far away from equilibrium is a challenging task. We discuss and experimentally confirm a series expansion for a driven classical system, relating the different non-equilibrium cumulants of the  observable conjugate to the driving protocol. This series is valid from micro- to macroscopic length scales, and it encompasses the fluctuation dissipation theorem. We apply it in experiments of a Brownian probe particle confined and driven by an optical potential and suspended in a nonlinear and non-Markovian fluid. The expansion states that the form of FDT remains valid away from equilibrium for Gaussian observables, up to the order presented. We show that this expansion agrees with the expansion of a known fluctuation theorem up to an unresolved  difference regarding moments versus cumulants.  
\end{abstract}

\maketitle

The fluctuation dissipation theorem (FDT)~\cite{callen_irreversibility_1951,kubo_fluctuation-dissipation_1966},  connecting response and fluctuations of equilibrium systems, is of fundamental importance for condensed matter, fluids, plasmas, or electromagnetic fields \cite{landau_statistical_1980,sitenko_electromagnetic_1967,rytov_principles_1987,hansen2013theory}.  
One of its remarkable properties is the validity at any length scale, be it the nanoscale, as for electric charges, or the macroscale, as for the macroscopic magnetization. 
It is however restricted to the linear regime, i.e., to situations close to equilibrium. Most previous research has been largely devoted to determining similar relations for non-equilibrium steady states~\cite{cugliandolo_fluctuation-dissipation_1997,ruelle_general_1998,crisanti_violation_2003,harada_equality_2005,speck_restoring_2006,deutsch_energy_2006,chetrite_fluctuation_2008,marconi_fluctuationdissipation_2008,saito_energy_2008,baiesi_fluctuations_2009,prost_generalized_2009,harada_macroscopic_2009,kruger_fluctuation_2009,seifert_fluctuation-dissipation_2010,baiesi_modified_2011,cugliandolo_effective_2011,verley_modified_2011,altaner_fluctuation-dissipation_2016,lippiello_nonequilibrium_2014,wu_generalized_2020,caprini_generalized_2021,baldovin_many_2022,johnsrud_generalized_2024} as well as for nonlinear responses~\cite{kubo_statistical-mechanical_1957,bernard_irreversible_1959,efremov_fluctuation_1968,bochkov_nonlinear_1981,stratonovich_nonlinear_1992,evans_statistical_2008,fuchs_integration_2005,holsten_thermodynamic_2021,oppenheim_nonlinear_1989,bouchaud_nonlinear_2005,lippiello_nonlinear_2008,andrieux_quantum_2008,lucarini_beyond_2012,diezemann_nonlinear_2012,kubo_statistical-mechanical_1957,wang_generalized_2002,andrieux_fluctuation_2007,colangeli_meaningful_2011,basu_frenetic_2015,basu_extrapolation_2018,maes_response_2020,caspers_nonlinear_2024, helden_measurement_2016}. A typical observation in the found relations is the explicit appearance of microscopic details -- sometimes referred to as frenetic components~\cite{maes_second_2014,maes_frenesy_2020}, or referred to as the information on the specific rule governing the time evolution \cite{lippiello_nonlinear_2008}-- often hampering a model independent formulation as well as systematic change of length scales such as coarse graining to macroscopic scales~\cite{colangeli_meaningful_2011,basu_frenetic_2015,maes_response_2020}. As a consequence, experimental test and application of such relations has indeed been successful for systems with a small number of accessible Markovian degrees of freedom~\cite{gomez-solano_experimental_2009,blickle_einstein_2007,mehl_experimental_2010,gomez-solano_fluctuations_2011,helden_measurement_2016}, for which the dynamics can be modeled.

In a different spirit, nonlinear fluctuation dissipation relations~\cite{bernard_irreversible_1959,efremov_fluctuation_1968,bochkov_general_1977,bochkov_nonlinear_1981,stratonovich_nonlinear_1992} and fluctuation theorems~\cite{bochkov_nonlinear_1981,jarzynski_nonequilibrium_1997,crooks_entropy_1999,andrieux_quantum_2008,seifert_stochastic_2012} have  been found, which can often be applied in absence of a specific model. But they have to our knowledge not been used to quantify the error of FDT.

In this manuscript, we discuss and experimentally confirm a series expansion for a driven classical system which  relates the different non-equilibrium cumulants of the observable conjugate to the driving protocol, up to a certain order in driving velocity. This series  (i) is valid from micro- to macroscopic length scales, (ii) it is  model independent,  and (iii) it encompasses the fluctuation dissipation theorem. We apply it in an experimental many body system of a Brownian probe particle interacting with worm-like micelles  and confined and driven by an optical potential. In these experiments, we demonstrate that the equilibrium third force cumulant quantifies the deviation from the fluctuation dissipation theorem in second order in driving. Notably, our theoretical predictions demonstrate that the form of the FDT remains valid for purely Gaussian observables within the displayed order.

Consider a classical system of stochastic degrees $\degr_t$ at time $t$, (weakly) coupled to a heat bath at temperature $T$. The system's potential energy $U$ depends on $\degr_t$ and on a time dependent deterministic protocol $\prot_t$, i.e., $U(\degr_t,\prot_t)$. The system is prepared in equilibrium at time $t \to-\infty$, with protocol value  $\prot_{-\infty}=\prot_t$, for simplicity~\cite{caspers_companion}. The time dependence of the protocol  drives the system away from equilibrium.

The derivative of $U$ with respect to $\trap_t$,   $F_t  \coloneqq \partial_{\prot_t} U(\degr_t,\trap_t)$ is the observable conjugate to $X_t$. E.g., if $X_t$ is a position as in our experiments, $F_t$ is (minus) the corresponding force. $F_t$ can be micro- or macroscopic; for example, let $\prot$ couple linearly to an observable $A(\degr)$, $U(\degr_t,\prot_t)=U(\degr_t,0)-\prot_t A(\degr_t)$, i.e., $F_t=  - A(\degr_t)$. Thus, if $A(\degr)$ is a macroscopic field, such as the macroscopic magnetization, $F_t$ is macroscopic. If $A(\degr)$ is the position of a molecular particle, $F_t$ is microscopic. The following remains valid if $X_t$ enters $U$  nonlinearly.

The statistical properties of $F_t$ in this non-equilibrium  situation are encoded in its cumulants and its correlations with another state observable $B_t=B(\degr_t,X_t)$, which we aim to study here. The well known fluctuation dissipation theorem connects the covariance in the unperturbed system and the first moment of $B$ under weak driving~\cite{callen_irreversibility_1951,kubo_fluctuation-dissipation_1966},
\begin{align}
\beta^2 \int_{-\infty}^t \mathrm{d}s\, \dot{\trap}_s  \langle B_t;F_s\rangle_\mathrm{eq} = \beta \left[\langle B_t\rangle  - \langle B_t \rangle_\mathrm{eq}\right]+ \mathcal{O}(\dot{\trap}^2),\label{eq:FDT}
\end{align}
with $\beta = 1/k_B T$ and Boltzmann constant $k_B$. $\langle \dots ;\dots \rangle$ denotes the 2nd cumulant, and similarly for higher orders below. 
Expectation values $\langle \dots \rangle_\mathrm{eq}$ are evaluated using equilibrium trajectories with protocol value fixed at $X_t$. Expectation values $\langle \dots \rangle$ are measured in the driven system, i.e., under the given time dependent protocol \cite{caspers_companion}. 

In Ref.~\cite{caspers_companion} we derive  identities connecting the non-equilibrium cumulants of $F_t$ and $B_t$ to different orders. 
These give rise to the following series involving the mentioned cumulants~\cite{caspers_companion},
\begin{align}
\begin{split}
  &\beta^2 \int_{-\infty}^t \mathrm{d}s\, \dot{\trap}_s \, \langle B_t;F_s\rangle = \beta \left[\langle B_t\rangle  - \langle B_t \rangle_\mathrm{eq}\right]\\
     &+ \frac{\beta^3}{2}  \int_{-\infty}^t \mathrm{d}s \int_{-\infty}^t \mathrm{d}s'\, \dot{\trap}_s \dot{\trap}_{s'} \langle B_t;F_s;F_{s'}\rangle_\mathrm{}\\
     &- \frac{\beta^4}{6}  \int_{-\infty}^t \mathrm{d}s \int_{-\infty}^t \mathrm{d}s' \int_{-\infty}^t \mathrm{d}s''\, \dot{\trap}_s \dot{\trap}_{s'} \dot{\trap}_{s''} \\
&\quad \quad  \times \langle B_t;F_s;F_{s'};F_{s''}\rangle+ \mathcal{O}(\dot{\trap}^4) ,
\end{split}\label{eq:NoiseIdentity}
\end{align}
As presented in Ref.~\cite{caspers_companion}, this series expansion can also be obtained from a known fluctuation theorem~\cite{bochkov_nonlinear_1981,andrieux_quantum_2008}, albeit with an open question regarding cumulants versus moments.

Eq.~\eqref{eq:NoiseIdentity} is, as indicated, correct up to fourth order in driving $\dot\trap_t$ under the assumption of local detailed balance\footnotetext[4]{The expansion in Eq.~\eqref{eq:NoiseIdentity} suggests the  dimensionless expansion parameter $\beta F \int_{t-\tau}^t \mathrm{d}s\, \dot x_s$ with cumulant relaxation time $\tau$}~\cite{maes_local_2021,Note4}. 
Expanding Eq.~\eqref{eq:NoiseIdentity} to  first order yields FDT in Eq.~\eqref{eq:FDT}, so that it is included in Eq.~\eqref{eq:NoiseIdentity}.
To higher orders, first and second cumulants do not fulfill FDT, and Eq.~\eqref{eq:NoiseIdentity} quantifies their difference in terms  of third and fourth cumulants of $F$ and $B$. Notably, the second to fourth lines of Eq.~\eqref{eq:NoiseIdentity} vanish for purely Gaussian distributed $F$ and $B$, so that first and second cumulants obey FDT to the given non-equilibrium order. 
It is important to note that in Eq.~\eqref{eq:NoiseIdentity} the protocol $\dot X$ appears as prefactors as well as in the non-equilibrium cumulants themselves, i.e., the latter are evaluated under application of driving.
As Eq.~\eqref{eq:NoiseIdentity} only requires measurement of $F$ and $B$, we use the notion of {\it model free}.

\begin{figure}
\includegraphics{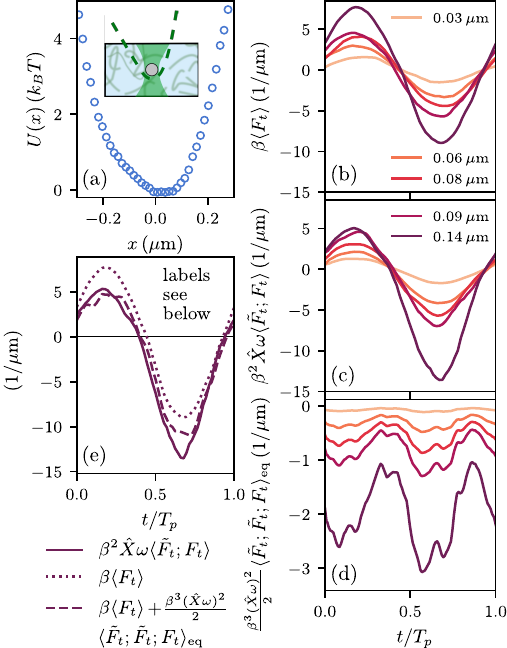 }
    \caption{(a) Asymmetric optical potential $U(\probe) = - k_B T \ln P(\probe)$ felt by the probe particle (inset sketch), with $P(\probe)$ the  probability distribution with the trap at rest. (b) Mean force $\beta \langle F_t\rangle$, (c) force covariance $\beta^2 \amp \omega \langle \Tilde{F}_t,F_t\rangle$, and (d) equilibrium third cumulant $\beta^2 (\amp\omega)^2 \langle \Tilde{F}_t;\Tilde{F}_t;F_t\rangle_\mathrm{eq}/2$, as functions of time, for driving  frequency $\omega = \SI{8.4}{\rad/\s}$ and amplitudes $\hat{X} = \{0.03,0.06,0.08,0.09,0.14\}\SI{}{\micro \m}$ as labeled. $T_p = \frac{2 \pi}{\omega}$. (e) Force covariance (solid line),  mean force (dotted line), and sum of mean force and third force cumulant (dashed line, Eq.~\eqref{eq:NoiseIdentityOscillatory}) for $\amp = \SI{0.14}{\micro\m}$.}
    \label{fig:TimeDependentForceCumulants}
\end{figure}

We exploit Eq.~\eqref{eq:NoiseIdentity} with experiments of Brownian particles interacting with micellar fluid. Specifically, we use silica particles of diameter $\sim \SI{1}{\micro \m}$ suspended in a $\SI{5}{\milli \Molar}$ equimolar solution of cetylpyridinium chloride monohydrate (CPyCl) and sodium salicylate (NaSal). At concentrations above the critical micellar concentration ($\gtrsim \SI{4}{\milli \Molar}$), this fluid is known to form giant worm-like micelles leading to a viscoelastic nonlinear behavior at ambient temperatures~\cite{cates_statics_1990}, see SM. At $\SI{5}{\milli \Molar}$, we determine the relaxation time of the fluid from microrheological recoil experiments, where a particle is first driven with a constant external force which is then suddenly removed, to be $\sim (3\pm\SI{0.2}){s}$~\cite{gomez-solano_transient_2015}.
A small amount of silica particles is added to the micellar solution which is contained in a rectangular capillary with $\SI{100}{\micro \m}$ height and kept at a temperature of $\SI{25}{\celsius}$. 
This sample is placed on a custom-built optical tweezer setup that uses a Gaussian laser beam of wavelength $\SI{532}{nm}$ and a 100$\times$ oil immersion objective (NA $=$ 1.45). 
 The laser beam yields a potential $U(x_t-X_t)$ as shown in Fig.~\ref{fig:TimeDependentForceCumulants}(a), centered at $X_t$, trapping one of the silica particles with coordinate $x_t$. $F_t = \partial_{X_t} U(x_t-X_t) $ is thus the force acting on the particle by the trapping potential (or vice versa). As the micellar degrees do not couple to $X$, they do not enter $F_t$ explicitly, and knowing how they enter $U$ is not required to apply Eq.~\eqref{eq:NoiseIdentity}. Thus, use of Eq.~\eqref{eq:NoiseIdentity} does not require detection of the positions of micellar particles, and applying it to such a complex fluid demonstrates its strength. 
 We consider $B\equiv F$, and made the potential asymmetric, to obtain a finite second order response of $\beta \langle F_t\rangle$. This allows to test Eq.~\eqref{eq:NoiseIdentity} to second order in our experiments, and it is achieved by a controlled lateral displacement of the vertically incident laser beam from the center of the objective lens (see SM). Notably, Eq.~\eqref{eq:NoiseIdentity} could also be tested in a purely viscous fluid,  which also shows a second order response due to the nonlinear potential. However, to demonstrate that   Eq.~\eqref{eq:NoiseIdentity} is valid beyond simple systems, we have chosen the more challenging case of a  micellar fluid.

To apply the driving protocol, the sample cell is moved, while the optical trap remains stationary in our experiments. This is achieved using a piezo-driven stage on which the sample is mounted and translated in an oscillating manner relative to the trap. In the fluid's rest frame, this yields a periodic motion of the potential minimum $X_t$, i.e., the  protocol, 
\begin{align}
\trap_t = \amp \sin(\omega t),\label{eq:prot}
\end{align}
with amplitude $\amp$ and frequency $\omega$.
Particle trajectories are recorded with a frame rate of $\sim \SI{150}{\Hz}$ using a video camera and particle positions are determined using a custom MATLAB algorithm.
To yield sufficient statistics, each protocol ($\amp$, $\omega$) was measured over 1400s. We allowed the system to reach a steady state by recording trajectories only after at least 5 oscillation periods have passed. Thereafter, no further equilibration was visible in the data.  
Prior to each non-equilibrium protocol, we  recorded particle trajectories for another 1000s with $X_t$ at rest. This equilibrium data were used to check that the form of $U$ does not vary between measurements, and also to  obtain the force cumulants under equilibrium conditions.

With the protocol of Eq.~\eqref{eq:prot}, Eq.~\eqref{eq:NoiseIdentity} takes, expanded to second order, the form 
\begin{align}
\begin{split}
    \beta^2 \amp \omega  \langle \tilde{F}_t;F_t\rangle &= \beta \langle F_t\rangle + \frac{ \beta^3 \amp^2\omega^2}{2}\langle \tilde{F}_t;\tilde{F}_t;F_t\rangle \\
    &\quad+ \mathcal{O}((\hat{X}\omega)^3),
    \end{split}\label{eq:NoiseIdentityOscillatory}
\end{align}
where the tilde denotes cosine transform, i.e., $\Tilde{F}_t \equiv \int_{-\infty}^t\mathrm{d}s \, \cos(\omega s )F_s$. We restrict the  analysis to the lowest nontrivial, i.e., second, order and expanded Eq.~\eqref{eq:NoiseIdentity} accordingly, also using $\langle F_t\rangle_{\mathrm{eq}}=0$. Notably, to extract the second order contribution from the last term in Eq.~\eqref{eq:NoiseIdentityOscillatory}, the third cumulant in Eq.~\eqref{eq:NoiseIdentityOscillatory} is replaced by its equilibrium version. This is because of the prefactor $\amp^2\omega^2$, and will be done in the following analysis.

The cumulants in  Eq.~\eqref{eq:NoiseIdentityOscillatory} depend on time $t$ in a periodic manner, as shown in   Figs.~\ref{fig:TimeDependentForceCumulants}(b)-(d), for $\omega = \SI{8.4}{\rad/\s}$\footnotetext[2]{$\omega$ is determined from the power spectral density, thus carrying an error depending on the length of the measurement.}~\cite{Note2} and driving amplitudes ranging from $\amp = \SI{0.03}{\micro \m}$ (light) to $\amp = \SI{0.14}{\micro \m}$ (dark). Fig.~\ref{fig:TimeDependentForceCumulants}(b) shows the mean force, which, as expected for a driven oscillator, is a periodic function with period $T_p=\frac{2\pi}{\omega}$. For the smallest amplitude shown, the mean force is nearly harmonic with frequency $\omega$, as expected from linear response. With growing amplitude, higher harmonics occur, as expected from nonlinear response. This asymmetric system shows second order response with expected frequencies of $2\omega$ and $0\omega$.

Fig.~\ref{fig:TimeDependentForceCumulants}(c) shows the force covariance for the same parameters and color code. For small amplitude $\hat X$, the curves in Figs.~\ref{fig:TimeDependentForceCumulants}(b) and (c) are equal within experimental accuracy, as analyzed in detail below. For larger driving amplitude, the force covariance develops higher harmonics with signatures of second order. Very little is known  about the properties of such non-equilibrium fluctuations, and quantifying these is difficult. It is notable that the curves in Fig.~\ref{fig:TimeDependentForceCumulants}(c), for larger amplitudes, deviate from  Fig.~\ref{fig:TimeDependentForceCumulants}(b), the deviation which we claim to be quantified by Eq.~\eqref{eq:NoiseIdentityOscillatory}.

Fig.~\ref{fig:TimeDependentForceCumulants}(d) shows the third cumulant of force for the same parameters and color code. We have here restricted to the equilibrium cumulant as it appears in Eq.~\eqref{eq:NoiseIdentityOscillatory}, multiplied by $(\hat X\omega)^2$. The curves in Fig.~\ref{fig:TimeDependentForceCumulants}(d) thus  differ only because of the factor $(\hat X\omega)^2$. They thus scale quadratically in driving velocity and only show frequencies of  $2\omega$ and $0\omega$.  

Eq.~\eqref{eq:NoiseIdentityOscillatory} states that, in the shown range of amplitudes, the curves in Fig.~\ref{fig:TimeDependentForceCumulants}(c) are given by the sum of the curves in Figs.~\ref{fig:TimeDependentForceCumulants}(b) and (d).
For $\hat{X} = \SI{0.14}{\micro \m}$ the respective summed curve is shown as a dashed line together with the mean force and the force covariance in Fig.~\ref{fig:TimeDependentForceCumulants}(e). The agreement is convincing and a  confirmation of Eq.~\eqref{eq:NoiseIdentityOscillatory}. 

To test this prediction systematically, we dissect the curves in Figs.~\ref{fig:TimeDependentForceCumulants} (b), (c) and (d) into the contributions from harmonics with frequencies $0\omega$, $\omega$, and $2\omega$, respectively, i.e., we expand the cumulant of order $n$ into harmonics with frequency $m\omega$, 
\begin{align}
\begin{split}
 \beta^n (\amp \omega)^{n-1}\langle (\Tilde{F}_t;)^{n-1} F_t\rangle &= \sum_{m=0}^\infty A_m^{(n)} \sin( m\omega t +\phi_m^{(n)}),\label{eq:harm}
    \end{split}
\end{align}
where the coefficients $A_m^{(n)}$ depend on  $\amp\omega$.
We set $\phi_0^{(n)}\equiv \pi/2$ for consistency.
\begin{figure}
    \centering
  \includegraphics{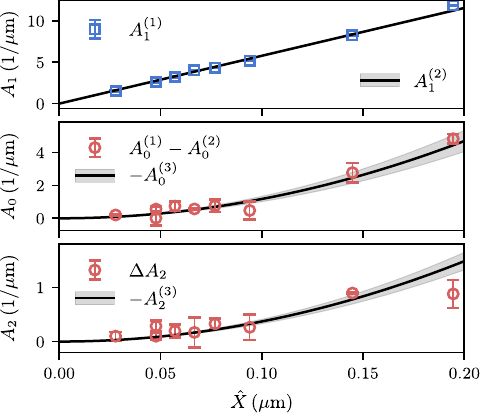}
    \caption{Coefficients $A_m$ corresponding to harmonics with frequency $m\omega$, as a function of  driving amplitude $\hat{X}$,  for  $\omega = \SI{8.4}{\rad/\s}$.
    Top panel: $m=1$, demonstrating   FDT.  Lower panels: Difference of first and second cumulants (data points), and third cumulant (lines) for $m=0$ and $m=2$. The agreement confirms Eq.~\eqref{eq:NoiseIdentityOscillatory}.
    $\Delta A_2 \equiv [(A_2^{(1)} )^2+(A_2^{(2)})^2-2 A_2^{(1)} A_2^{(2)}\cos(\phi_2^{(1)}-\phi_2^{(2)})]^{1/2}$ \cite{Note3}.
    Error bars and bands are obtained from partitioning trajectories into two pieces.}
    \label{fig:amplitude}
\end{figure}
Eq.~\eqref{eq:NoiseIdentityOscillatory}, projected on the harmonic of order $m$, yields relations between coefficients and phases for each $m$, which we can test.

Fig.~\ref{fig:amplitude} shows 
the coefficients $A_m^{(n)}$ as a function of driving amplitude $\hat{X}$. The top panel gives the order $m=1$, which is seen to be linear in $\hat X$ for the range shown, as expected from linear response. The graph shows the mean force (data points) as well as the force covariance (line). The latter is evaluated from equilibrium trajectories. 
The agreement in this panel, for the range of sufficiently small $\hat{X}$, is expected from the fluctuation dissipation theorem. 

The center and lower panels in Fig.~\ref{fig:amplitude} show the orders $m=0$ and $m=2$, respectively. Specifically, these panels present the difference of first and second cumulants in Eq.~\eqref{eq:NoiseIdentityOscillatory} (data points), i.e., $\beta \langle F_t\rangle - \beta^2 \hat{X}\omega \langle \tilde{F}_t;F_t\rangle$, together with the third cumulant (line), $-\beta^3 (\hat{X}\omega)^2 \langle \Tilde{F}_t;\Tilde{F}_t;F_t\rangle_\mathrm{eq}/2$, evaluated from equilibrium measurements \footnotetext[3]{
As the phases $\phi_2^{(1)}$ and $\phi_2^{(2)}$ may differ, the coefficient $\Delta A_2$ of the difference of first and second cumulants is found via $\Delta A_2 \equiv \sqrt{\left(A_2^{(1)} \right)^2+\left(A_2^{(2)}\right)^2-2 A_2^{(1)} A_2^{(2)}\cos\left(\phi_2^{(1)}-\phi_2^{(2)}\right)}$. For $m=0$, i.e., the zero frequency contribution, there is no phase by definition, and the coefficients $A_0^{(n)}$ can be compared directly}~\cite{Note3}.
The latter is shown as a parabola with curvature obtained from the third force cumulant at equilibrium.
The data points in this graph thus quantify the deviation from FDT, with the line giving the prediction of Eq.~\eqref{eq:NoiseIdentityOscillatory} for this deviation. 
The agreement is convincing for both $m=0$ and $m=2$,
supporting the validity of Eq.~\eqref{eq:NoiseIdentityOscillatory}.

\begin{figure}
    \centering
 \includegraphics{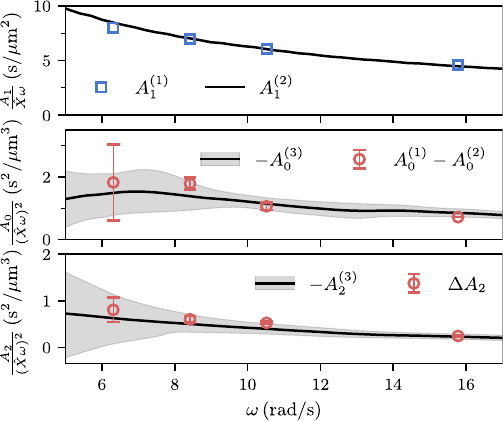}
    \caption{Coefficients $A_m$, normalized as labeled, as a function of frequency $\omega$. The agreement in the top panel confirms FDT, the agreement in the lower panels confirms Eq.~\eqref{eq:NoiseIdentityOscillatory}. Each data point is obtained from averaging over driving amplitudes taking the respective scaling of $A_m$ with $\hat{X}$ into account (see SM). Error bars are obtained from partitioning trajectories into two pieces (data points) and from the standard deviation between separate series of measurements (grey area).}
    \label{fig:omega}
\end{figure}

As the data in the top panel of Fig.~\ref{fig:amplitude} grow linearly and the ones in the center and lower panels grow quadratically with $\hat X$, we fit a line and a parabola to  obtain the respective slope and curvature for each $m$. 
The obtained values -- divided by the respective power in $\omega$ -- are shown in Fig.~\ref{fig:omega} as a function of frequency $\omega$. 
We observe convincing agreement for the measured frequencies further suporting Eq.~\eqref{eq:NoiseIdentityOscillatory}.
In tendency, the coefficients decrease with increasing $\omega$. 
Noteably, the statistical accuracy of the data points decreases with decreasing $\omega$; With smaller $\omega$, longer trajectories are required, especially in a fluid with pronounced memory, as the period of the cosine transform increases. 

\begin{figure}
\includegraphics{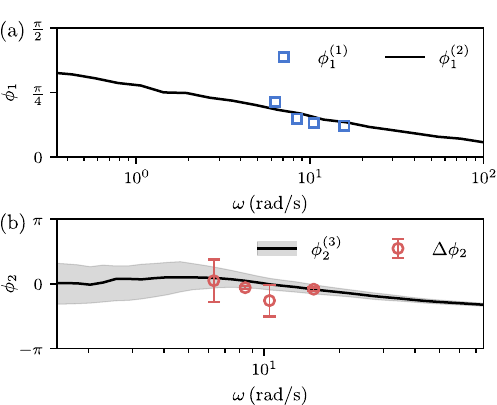}
    \caption{Phase angles $\phi_1$ and $\phi_2$ of the harmonics in Eq.~\eqref{eq:harm}, as functions of $\omega$. Top panel shows phases of linear response. Lower panel shows the phase of second order response, comparing the contributions of the terms in Eq.~\eqref{eq:NoiseIdentityOscillatory}. $\Delta \phi_2 \equiv \arctan \frac{A_2^{(1)}\sin \phi_2^{(1)}-A_2^{(2)}\sin \phi_2^{(2)}}{A_2^{(1)}\cos \phi_2^{(1)}-A_2^{(2)}\cos \phi_2^{(2)}}$, \cite{Note3}. Error bars are obtained from partitioning trajectories into two pieces. Grey error band is obtained as the standard deviation between separate series of measurements.}
    \label{fig:phasesaveraged}
\end{figure}

Fig.~\ref{fig:phasesaveraged} provides the final test of Eq.~\eqref{eq:NoiseIdentityOscillatory}, namely the phases $\phi_m^{(n)}$ of Eq.~\eqref{eq:harm}. These hardly depend on driving amplitude and the shown data are averaged over the measured values of $\hat X$. The top panel of Fig.~\ref{fig:phasesaveraged} shows the order $m=1$, i.e., the linear response, with convincing agreement. The phase angle for $m=1$ is small for the frequencies measured, indicating that the force $F_t$ is almost in phase with the protocol $X_t$, as for an elastic material. The black curve, extracted from equilibrium data, shows that with smaller $\omega$ the phase increases, presumably reaching $\pi/2$ in the limit of $\omega\to0$.
The slow increase with decreasing $\omega$ displays the slow nature of the investigated system.

The lower panel shows the phase for $m=2$, i.e., to second order. While the agreement between the line and the data confirms  Eq.~\eqref{eq:NoiseIdentityOscillatory}, the graph shows that the difference of phases of first and second cumulants are rather small. In other words, first and second cumulants deviate noticeable in amplitude, seen in Fig.~\ref{fig:amplitude}, but not so much in phase.


We presented and tested a non-equilibrium fluctuation expansion for a driven classical system, emphasizing the validity on various length scales. Indeed, such relations are necessary, e.g., for a systematic coarse graining of non-equilibrium systems. The identity is confirmed for experiments of a Brownian particle interacting with a complex surrounding. 
Future work can explore other systems and aim to clarify the relation to the mentioned fluctuation theorems \cite{bochkov_nonlinear_1981,andrieux_quantum_2008}. It is also important to investigate use of Eq.~\eqref{eq:NoiseIdentity} for treating systems far away from equilibrium.  

This project was funded by the Deutsche Forschungsgemeinschaft (DFG), Grant No.~SFB 1432 (Project ID 425217212)—Project C05.


\begin{thebibliography}{68}%
\makeatletter
\providecommand \@ifxundefined [1]{%
 \@ifx{#1\undefined}
}%
\providecommand \@ifnum [1]{%
 \ifnum #1\expandafter \@firstoftwo
 \else \expandafter \@secondoftwo
 \fi
}%
\providecommand \@ifx [1]{%
 \ifx #1\expandafter \@firstoftwo
 \else \expandafter \@secondoftwo
 \fi
}%
\providecommand \natexlab [1]{#1}%
\providecommand \enquote  [1]{``#1''}%
\providecommand \bibnamefont  [1]{#1}%
\providecommand \bibfnamefont [1]{#1}%
\providecommand \citenamefont [1]{#1}%
\providecommand \href@noop [0]{\@secondoftwo}%
\providecommand \href [0]{\begingroup \@sanitize@url \@href}%
\providecommand \@href[1]{\@@startlink{#1}\@@href}%
\providecommand \@@href[1]{\endgroup#1\@@endlink}%
\providecommand \@sanitize@url [0]{\catcode `\\12\catcode `\$12\catcode
  `\&12\catcode `\#12\catcode `\^12\catcode `\_12\catcode `\%12\relax}%
\providecommand \@@startlink[1]{}%
\providecommand \@@endlink[0]{}%
\providecommand \url  [0]{\begingroup\@sanitize@url \@url }%
\providecommand \@url [1]{\endgroup\@href {#1}{\urlprefix }}%
\providecommand \urlprefix  [0]{URL }%
\providecommand \Eprint [0]{\href }%
\providecommand \doibase [0]{https://doi.org/}%
\providecommand \selectlanguage [0]{\@gobble}%
\providecommand \bibinfo  [0]{\@secondoftwo}%
\providecommand \bibfield  [0]{\@secondoftwo}%
\providecommand \translation [1]{[#1]}%
\providecommand \BibitemOpen [0]{}%
\providecommand \bibitemStop [0]{}%
\providecommand \bibitemNoStop [0]{.\EOS\space}%
\providecommand \EOS [0]{\spacefactor3000\relax}%
\providecommand \BibitemShut  [1]{\csname bibitem#1\endcsname}%
\let\auto@bib@innerbib\@empty
\bibitem [{\citenamefont {Callen}\ and\ \citenamefont
  {Welton}(1951)}]{callen_irreversibility_1951}%
  \BibitemOpen
  \bibfield  {author} {\bibinfo {author} {\bibfnamefont {H.~B.}\ \bibnamefont
  {Callen}}\ and\ \bibinfo {author} {\bibfnamefont {T.~A.}\ \bibnamefont
  {Welton}},\ }\href {https://doi.org/10.1103/PhysRev.83.34} {\bibfield
  {journal} {\bibinfo  {journal} {Phys. Rev.}\ }\textbf {\bibinfo {volume}
  {83}},\ \bibinfo {pages} {34} (\bibinfo {year} {1951})}\BibitemShut {NoStop}%
\bibitem [{\citenamefont {Kubo}(1966)}]{kubo_fluctuation-dissipation_1966}%
  \BibitemOpen
  \bibfield  {author} {\bibinfo {author} {\bibfnamefont {R.}~\bibnamefont
  {Kubo}},\ }\href {https://doi.org/10.1088/0034-4885/29/1/306} {\bibfield
  {journal} {\bibinfo  {journal} {Rep. Prog. Phys.}\ }\textbf {\bibinfo
  {volume} {29}},\ \bibinfo {pages} {255} (\bibinfo {year} {1966})}\BibitemShut
  {NoStop}%
\bibitem [{\citenamefont {Landau}\ and\ \citenamefont
  {Lifshitz}(1980)}]{landau_statistical_1980}%
  \BibitemOpen
  \bibfield  {author} {\bibinfo {author} {\bibfnamefont {L.~D.}\ \bibnamefont
  {Landau}}\ and\ \bibinfo {author} {\bibfnamefont {E.~M.}\ \bibnamefont
  {Lifshitz}},\ }\href@noop {} {\emph {\bibinfo {title} {Statistical
  {Physics}}}}\ (\bibinfo  {publisher} {Pergamon},\ \bibinfo {year}
  {1980})\BibitemShut {NoStop}%
\bibitem [{\citenamefont {Sitenko}(1967)}]{sitenko_electromagnetic_1967}%
  \BibitemOpen
  \bibfield  {author} {\bibinfo {author} {\bibfnamefont {A.~G.}\ \bibnamefont
  {Sitenko}},\ }\href@noop {} {\emph {\bibinfo {title} {Electromagnetic
  {Fluctuations} {In} {Plasma}}}}\ (\bibinfo  {publisher} {Academic {Press},
  {New York}},\ \bibinfo {year} {1967})\BibitemShut {NoStop}%
\bibitem [{\citenamefont {Rytov}\ \emph {et~al.}(1987)\citenamefont {Rytov},
  \citenamefont {Kravtsov},\ and\ \citenamefont
  {Tatarskii}}]{rytov_principles_1987}%
  \BibitemOpen
  \bibfield  {author} {\bibinfo {author} {\bibfnamefont {S.~M.}\ \bibnamefont
  {Rytov}}, \bibinfo {author} {\bibfnamefont {I.~A.}\ \bibnamefont
  {Kravtsov}},\ and\ \bibinfo {author} {\bibfnamefont {V.}~\bibnamefont
  {Tatarskii}},\ }\href@noop {} {\emph {\bibinfo {title} {Principles of
  {Statistical} {Radiophysics}}}}\ (\bibinfo  {publisher} {Springer-Verlag},\
  \bibinfo {year} {1987})\BibitemShut {NoStop}%
\bibitem [{\citenamefont {Hansen}\ and\ \citenamefont
  {McDonald}(2013)}]{hansen2013theory}%
  \BibitemOpen
  \bibfield  {author} {\bibinfo {author} {\bibfnamefont {J.~P.}\ \bibnamefont
  {Hansen}}\ and\ \bibinfo {author} {\bibfnamefont {I.~R.}\ \bibnamefont
  {McDonald}},\ }\href {https://books.google.co.jp/books?id=pbJfOUqZVSgC}
  {\emph {\bibinfo {title} {Theory of Simple Liquids: with Applications to Soft
  Matter}}}\ (\bibinfo  {publisher} {Elsevier Science},\ \bibinfo {year}
  {2013})\BibitemShut {NoStop}%
\bibitem [{\citenamefont {Cugliandolo}\ \emph {et~al.}(1997)\citenamefont
  {Cugliandolo}, \citenamefont {Dean},\ and\ \citenamefont
  {Kurchan}}]{cugliandolo_fluctuation-dissipation_1997}%
  \BibitemOpen
  \bibfield  {author} {\bibinfo {author} {\bibfnamefont {L.~F.}\ \bibnamefont
  {Cugliandolo}}, \bibinfo {author} {\bibfnamefont {D.~S.}\ \bibnamefont
  {Dean}},\ and\ \bibinfo {author} {\bibfnamefont {J.}~\bibnamefont
  {Kurchan}},\ }\href {https://doi.org/10.1103/PhysRevLett.79.2168} {\bibfield
  {journal} {\bibinfo  {journal} {Phys. Rev. Lett.}\ }\textbf {\bibinfo
  {volume} {79}},\ \bibinfo {pages} {2168} (\bibinfo {year}
  {1997})}\BibitemShut {NoStop}%
\bibitem [{\citenamefont {Ruelle}(1998)}]{ruelle_general_1998}%
  \BibitemOpen
  \bibfield  {author} {\bibinfo {author} {\bibfnamefont {D.}~\bibnamefont
  {Ruelle}},\ }\href {https://doi.org/10.1016/S0375-9601(98)00419-8} {\bibfield
   {journal} {\bibinfo  {journal} {Phys. Lett. A}\ }\textbf {\bibinfo {volume}
  {245}},\ \bibinfo {pages} {220} (\bibinfo {year} {1998})}\BibitemShut
  {NoStop}%
\bibitem [{\citenamefont {Crisanti}\ and\ \citenamefont
  {Ritort}(2003)}]{crisanti_violation_2003}%
  \BibitemOpen
  \bibfield  {author} {\bibinfo {author} {\bibfnamefont {A.}~\bibnamefont
  {Crisanti}}\ and\ \bibinfo {author} {\bibfnamefont {F.}~\bibnamefont
  {Ritort}},\ }\href {https://doi.org/10.1088/0305-4470/36/21/201} {\bibfield
  {journal} {\bibinfo  {journal} {J. Phys. A: Math. Gen.}\ }\textbf {\bibinfo
  {volume} {36}},\ \bibinfo {pages} {R181} (\bibinfo {year}
  {2003})}\BibitemShut {NoStop}%
\bibitem [{\citenamefont {Harada}\ and\ \citenamefont
  {Sasa}(2005)}]{harada_equality_2005}%
  \BibitemOpen
  \bibfield  {author} {\bibinfo {author} {\bibfnamefont {T.}~\bibnamefont
  {Harada}}\ and\ \bibinfo {author} {\bibfnamefont {S.-i.}\ \bibnamefont
  {Sasa}},\ }\href {https://doi.org/10.1103/PhysRevLett.95.130602} {\bibfield
  {journal} {\bibinfo  {journal} {Phys. Rev. Lett.}\ }\textbf {\bibinfo
  {volume} {95}},\ \bibinfo {pages} {130602} (\bibinfo {year}
  {2005})}\BibitemShut {NoStop}%
\bibitem [{\citenamefont {Speck}\ and\ \citenamefont
  {Seifert}(2006)}]{speck_restoring_2006}%
  \BibitemOpen
  \bibfield  {author} {\bibinfo {author} {\bibfnamefont {T.}~\bibnamefont
  {Speck}}\ and\ \bibinfo {author} {\bibfnamefont {U.}~\bibnamefont
  {Seifert}},\ }\href {https://doi.org/10.1209/epl/i2005-10549-4} {\bibfield
  {journal} {\bibinfo  {journal} {EPL}\ }\textbf {\bibinfo {volume} {74}},\
  \bibinfo {pages} {391} (\bibinfo {year} {2006})}\BibitemShut {NoStop}%
\bibitem [{\citenamefont {Deutsch}\ and\ \citenamefont
  {Narayan}(2006)}]{deutsch_energy_2006}%
  \BibitemOpen
  \bibfield  {author} {\bibinfo {author} {\bibfnamefont {J.~M.}\ \bibnamefont
  {Deutsch}}\ and\ \bibinfo {author} {\bibfnamefont {O.}~\bibnamefont
  {Narayan}},\ }\href {https://doi.org/10.1103/PhysRevE.74.026112} {\bibfield
  {journal} {\bibinfo  {journal} {Phys. Rev. E}\ }\textbf {\bibinfo {volume}
  {74}},\ \bibinfo {pages} {026112} (\bibinfo {year} {2006})}\BibitemShut
  {NoStop}%
\bibitem [{\citenamefont {Chetrite}\ \emph {et~al.}(2008)\citenamefont
  {Chetrite}, \citenamefont {Falkovich},\ and\ \citenamefont
  {Gawedzki}}]{chetrite_fluctuation_2008}%
  \BibitemOpen
  \bibfield  {author} {\bibinfo {author} {\bibfnamefont {R.}~\bibnamefont
  {Chetrite}}, \bibinfo {author} {\bibfnamefont {G.}~\bibnamefont
  {Falkovich}},\ and\ \bibinfo {author} {\bibfnamefont {K.}~\bibnamefont
  {Gawedzki}},\ }\href {https://doi.org/10.1088/1742-5468/2008/08/P08005}
  {\bibfield  {journal} {\bibinfo  {journal} {J. Stat. Mech.}\ }\textbf
  {\bibinfo {volume} {2008}},\ \bibinfo {pages} {P08005} (\bibinfo {year}
  {2008})}\BibitemShut {NoStop}%
\bibitem [{\citenamefont {Marconi}\ \emph {et~al.}(2008)\citenamefont
  {Marconi}, \citenamefont {Puglisi}, \citenamefont {Rondoni},\ and\
  \citenamefont {Vulpiani}}]{marconi_fluctuationdissipation_2008}%
  \BibitemOpen
  \bibfield  {author} {\bibinfo {author} {\bibfnamefont {U.~M.~B.}\
  \bibnamefont {Marconi}}, \bibinfo {author} {\bibfnamefont {A.}~\bibnamefont
  {Puglisi}}, \bibinfo {author} {\bibfnamefont {L.}~\bibnamefont {Rondoni}},\
  and\ \bibinfo {author} {\bibfnamefont {A.}~\bibnamefont {Vulpiani}},\ }\href
  {https://doi.org/10.1016/j.physrep.2008.02.002} {\bibfield  {journal}
  {\bibinfo  {journal} {Phys. Rep.}\ }\textbf {\bibinfo {volume} {461}},\
  \bibinfo {pages} {111} (\bibinfo {year} {2008})}\BibitemShut {NoStop}%
\bibitem [{\citenamefont {Saito}(2008)}]{saito_energy_2008}%
  \BibitemOpen
  \bibfield  {author} {\bibinfo {author} {\bibfnamefont {K.}~\bibnamefont
  {Saito}},\ }\href {https://doi.org/10.1209/0295-5075/83/50006} {\bibfield
  {journal} {\bibinfo  {journal} {Europhys. Lett.}\ }\textbf {\bibinfo {volume}
  {83}},\ \bibinfo {pages} {50006} (\bibinfo {year} {2008})}\BibitemShut
  {NoStop}%
\bibitem [{\citenamefont {Baiesi}\ \emph {et~al.}(2009)\citenamefont {Baiesi},
  \citenamefont {Maes},\ and\ \citenamefont
  {Wynants}}]{baiesi_fluctuations_2009}%
  \BibitemOpen
  \bibfield  {author} {\bibinfo {author} {\bibfnamefont {M.}~\bibnamefont
  {Baiesi}}, \bibinfo {author} {\bibfnamefont {C.}~\bibnamefont {Maes}},\ and\
  \bibinfo {author} {\bibfnamefont {B.}~\bibnamefont {Wynants}},\ }\href
  {https://doi.org/10.1103/PhysRevLett.103.010602} {\bibfield  {journal}
  {\bibinfo  {journal} {Phys. Rev. Lett.}\ }\textbf {\bibinfo {volume} {103}},\
  \bibinfo {pages} {010602} (\bibinfo {year} {2009})}\BibitemShut {NoStop}%
\bibitem [{\citenamefont {Prost}\ \emph {et~al.}(2009)\citenamefont {Prost},
  \citenamefont {Joanny},\ and\ \citenamefont
  {Parrondo}}]{prost_generalized_2009}%
  \BibitemOpen
  \bibfield  {author} {\bibinfo {author} {\bibfnamefont {J.}~\bibnamefont
  {Prost}}, \bibinfo {author} {\bibfnamefont {J.-F.}\ \bibnamefont {Joanny}},\
  and\ \bibinfo {author} {\bibfnamefont {J.~M.~R.}\ \bibnamefont {Parrondo}},\
  }\href {https://doi.org/10.1103/PhysRevLett.103.090601} {\bibfield  {journal}
  {\bibinfo  {journal} {Phys. Rev. Lett.}\ }\textbf {\bibinfo {volume} {103}},\
  \bibinfo {pages} {090601} (\bibinfo {year} {2009})}\BibitemShut {NoStop}%
\bibitem [{\citenamefont {Harada}(2009)}]{harada_macroscopic_2009}%
  \BibitemOpen
  \bibfield  {author} {\bibinfo {author} {\bibfnamefont {T.}~\bibnamefont
  {Harada}},\ }\href {https://doi.org/10.1103/PhysRevE.79.030106} {\bibfield
  {journal} {\bibinfo  {journal} {Phys. Rev. E}\ }\textbf {\bibinfo {volume}
  {79}},\ \bibinfo {pages} {030106} (\bibinfo {year} {2009})}\BibitemShut
  {NoStop}%
\bibitem [{\citenamefont {Krüger}\ and\ \citenamefont
  {Fuchs}(2009)}]{kruger_fluctuation_2009}%
  \BibitemOpen
  \bibfield  {author} {\bibinfo {author} {\bibfnamefont {M.}~\bibnamefont
  {Krüger}}\ and\ \bibinfo {author} {\bibfnamefont {M.}~\bibnamefont
  {Fuchs}},\ }\href {https://doi.org/10.1103/PhysRevLett.102.135701} {\bibfield
   {journal} {\bibinfo  {journal} {Phys. Rev. Lett.}\ }\textbf {\bibinfo
  {volume} {102}},\ \bibinfo {pages} {135701} (\bibinfo {year}
  {2009})}\BibitemShut {NoStop}%
\bibitem [{\citenamefont {Seifert}\ and\ \citenamefont
  {Speck}(2010)}]{seifert_fluctuation-dissipation_2010}%
  \BibitemOpen
  \bibfield  {author} {\bibinfo {author} {\bibfnamefont {U.}~\bibnamefont
  {Seifert}}\ and\ \bibinfo {author} {\bibfnamefont {T.}~\bibnamefont
  {Speck}},\ }\href {https://doi.org/10.1209/0295-5075/89/10007} {\bibfield
  {journal} {\bibinfo  {journal} {EPL}\ }\textbf {\bibinfo {volume} {89}},\
  \bibinfo {pages} {10007} (\bibinfo {year} {2010})}\BibitemShut {NoStop}%
\bibitem [{\citenamefont {Baiesi}\ \emph {et~al.}(2011)\citenamefont {Baiesi},
  \citenamefont {Maes},\ and\ \citenamefont {Wynants}}]{baiesi_modified_2011}%
  \BibitemOpen
  \bibfield  {author} {\bibinfo {author} {\bibfnamefont {M.}~\bibnamefont
  {Baiesi}}, \bibinfo {author} {\bibfnamefont {C.}~\bibnamefont {Maes}},\ and\
  \bibinfo {author} {\bibfnamefont {B.}~\bibnamefont {Wynants}},\ }\href
  {https://doi.org/10.1098/rspa.2011.0046} {\bibfield  {journal} {\bibinfo
  {journal} {Proc. R. Soc. A: Math. Phys. Eng. Sci.}\ }\textbf {\bibinfo
  {volume} {467}},\ \bibinfo {pages} {2792} (\bibinfo {year}
  {2011})}\BibitemShut {NoStop}%
\bibitem [{\citenamefont {Cugliandolo}(2011)}]{cugliandolo_effective_2011}%
  \BibitemOpen
  \bibfield  {author} {\bibinfo {author} {\bibfnamefont {L.~F.}\ \bibnamefont
  {Cugliandolo}},\ }\href {https://doi.org/10.1088/1751-8113/44/48/483001}
  {\bibfield  {journal} {\bibinfo  {journal} {J. Phys. A: Math. Theor.}\
  }\textbf {\bibinfo {volume} {44}},\ \bibinfo {pages} {483001} (\bibinfo
  {year} {2011})}\BibitemShut {NoStop}%
\bibitem [{\citenamefont {Verley}\ \emph {et~al.}(2011)\citenamefont {Verley},
  \citenamefont {Chétrite},\ and\ \citenamefont
  {Lacoste}}]{verley_modified_2011}%
  \BibitemOpen
  \bibfield  {author} {\bibinfo {author} {\bibfnamefont {G.}~\bibnamefont
  {Verley}}, \bibinfo {author} {\bibfnamefont {R.}~\bibnamefont {Chétrite}},\
  and\ \bibinfo {author} {\bibfnamefont {D.}~\bibnamefont {Lacoste}},\ }\href
  {https://doi.org/10.1088/1742-5468/2011/10/P10025} {\bibfield  {journal}
  {\bibinfo  {journal} {J. Stat. Mech.}\ }\textbf {\bibinfo {volume} {2011}},\
  \bibinfo {pages} {10025} (\bibinfo {year} {2011})}\BibitemShut {NoStop}%
\bibitem [{\citenamefont {Altaner}\ \emph {et~al.}(2016)\citenamefont
  {Altaner}, \citenamefont {Polettini},\ and\ \citenamefont
  {Esposito}}]{altaner_fluctuation-dissipation_2016}%
  \BibitemOpen
  \bibfield  {author} {\bibinfo {author} {\bibfnamefont {B.}~\bibnamefont
  {Altaner}}, \bibinfo {author} {\bibfnamefont {M.}~\bibnamefont {Polettini}},\
  and\ \bibinfo {author} {\bibfnamefont {M.}~\bibnamefont {Esposito}},\ }\href
  {https://doi.org/10.1103/PhysRevLett.117.180601} {\bibfield  {journal}
  {\bibinfo  {journal} {Phys. Rev. Lett.}\ }\textbf {\bibinfo {volume} {117}},\
  \bibinfo {pages} {180601} (\bibinfo {year} {2016})}\BibitemShut {NoStop}%
\bibitem [{\citenamefont {Lippiello}\ \emph {et~al.}(2014)\citenamefont
  {Lippiello}, \citenamefont {Baiesi},\ and\ \citenamefont
  {Sarracino}}]{lippiello_nonequilibrium_2014}%
  \BibitemOpen
  \bibfield  {author} {\bibinfo {author} {\bibfnamefont {E.}~\bibnamefont
  {Lippiello}}, \bibinfo {author} {\bibfnamefont {M.}~\bibnamefont {Baiesi}},\
  and\ \bibinfo {author} {\bibfnamefont {A.}~\bibnamefont {Sarracino}},\ }\href
  {https://doi.org/10.1103/PhysRevLett.112.140602} {\bibfield  {journal}
  {\bibinfo  {journal} {Phys. Rev. Lett.}\ }\textbf {\bibinfo {volume} {112}},\
  \bibinfo {pages} {140602} (\bibinfo {year} {2014})}\BibitemShut {NoStop}%
\bibitem [{\citenamefont {Wu}\ and\ \citenamefont
  {Wang}(2020)}]{wu_generalized_2020}%
  \BibitemOpen
  \bibfield  {author} {\bibinfo {author} {\bibfnamefont {W.}~\bibnamefont
  {Wu}}\ and\ \bibinfo {author} {\bibfnamefont {J.}~\bibnamefont {Wang}},\
  }\href
  {https://www.frontiersin.org/journals/physics/articles/10.3389/fphy.2020.567523/full}
  {\bibfield  {journal} {\bibinfo  {journal} {Front. Phys.}\ }\textbf {\bibinfo
  {volume} {8}} (\bibinfo {year} {2020})}\BibitemShut {NoStop}%
\bibitem [{\citenamefont {Caprini}(2021)}]{caprini_generalized_2021}%
  \BibitemOpen
  \bibfield  {author} {\bibinfo {author} {\bibfnamefont {L.}~\bibnamefont
  {Caprini}},\ }\href {https://doi.org/10.1088/1742-5468/abffd4} {\bibfield
  {journal} {\bibinfo  {journal} {J. Stat. Mech.}\ }\textbf {\bibinfo {volume}
  {2021}},\ \bibinfo {pages} {063202} (\bibinfo {year} {2021})}\BibitemShut
  {NoStop}%
\bibitem [{\citenamefont {Baldovin}\ \emph {et~al.}(2022)\citenamefont
  {Baldovin}, \citenamefont {Caprini}, \citenamefont {Puglisi}, \citenamefont
  {Sarracino},\ and\ \citenamefont {Vulpiani}}]{baldovin_many_2022}%
  \BibitemOpen
  \bibfield  {author} {\bibinfo {author} {\bibfnamefont {M.}~\bibnamefont
  {Baldovin}}, \bibinfo {author} {\bibfnamefont {L.}~\bibnamefont {Caprini}},
  \bibinfo {author} {\bibfnamefont {A.}~\bibnamefont {Puglisi}}, \bibinfo
  {author} {\bibfnamefont {A.}~\bibnamefont {Sarracino}},\ and\ \bibinfo
  {author} {\bibfnamefont {A.}~\bibnamefont {Vulpiani}},\ }in\ \href
  {https://doi.org/10.1007/978-3-031-04458-8_3} {\emph {\bibinfo {booktitle}
  {Nonequilibrium {Thermodynamics} and {Fluctuation} {Kinetics}: {Modern}
  {Trends} and {Open} {Questions}}}}\ (\bibinfo  {publisher} {Springer
  International Publishing},\ \bibinfo {year} {2022})\BibitemShut {NoStop}%
\bibitem [{\citenamefont {Johnsrud}\ and\ \citenamefont
  {Golestanian}(2024)}]{johnsrud_generalized_2024}%
  \BibitemOpen
  \bibfield  {author} {\bibinfo {author} {\bibfnamefont {M.~K.}\ \bibnamefont
  {Johnsrud}}\ and\ \bibinfo {author} {\bibfnamefont {R.}~\bibnamefont
  {Golestanian}},\ }\href {https://doi.org/10.48550/arXiv.2409.14977} {\bibinfo
  {title} {Generalized {Fluctuation} {Dissipation} {Relations} for {Active}
  {Field} {Theories}}} (\bibinfo {year} {2024}),\ \bibinfo {note}
  {arXiv:2409.14977}\BibitemShut {NoStop}%
\bibitem [{\citenamefont {Kubo}(1957)}]{kubo_statistical-mechanical_1957}%
  \BibitemOpen
  \bibfield  {author} {\bibinfo {author} {\bibfnamefont {R.}~\bibnamefont
  {Kubo}},\ }\href {https://doi.org/10.1143/JPSJ.12.570} {\bibfield  {journal}
  {\bibinfo  {journal} {J. Phys. Soc. Jpn.}\ }\textbf {\bibinfo {volume}
  {12}},\ \bibinfo {pages} {570} (\bibinfo {year} {1957})}\BibitemShut
  {NoStop}%
\bibitem [{\citenamefont {Bernard}\ and\ \citenamefont
  {Callen}(1959)}]{bernard_irreversible_1959}%
  \BibitemOpen
  \bibfield  {author} {\bibinfo {author} {\bibfnamefont {W.}~\bibnamefont
  {Bernard}}\ and\ \bibinfo {author} {\bibfnamefont {H.~B.}\ \bibnamefont
  {Callen}},\ }\href {https://doi.org/10.1103/RevModPhys.31.1017} {\bibfield
  {journal} {\bibinfo  {journal} {Rev. Mod. Phys.}\ }\textbf {\bibinfo {volume}
  {31}},\ \bibinfo {pages} {1017} (\bibinfo {year} {1959})}\BibitemShut
  {NoStop}%
\bibitem [{\citenamefont {Efremov}(1968)}]{efremov_fluctuation_1968}%
  \BibitemOpen
  \bibfield  {author} {\bibinfo {author} {\bibfnamefont {G.~F.}\ \bibnamefont
  {Efremov}},\ }\href {http://www.jetp.ras.ru/cgi-bin/dn/e_028_06_1232.pdf}
  {\bibfield  {journal} {\bibinfo  {journal} {Zh. Eksp. Theor. Fiz.}\ }\textbf
  {\bibinfo {volume} {55}},\ \bibinfo {pages} {2322} (\bibinfo {year}
  {1968})}\BibitemShut {NoStop}%
\bibitem [{\citenamefont {Bochkov}\ and\ \citenamefont
  {Kuzovlev}(1981)}]{bochkov_nonlinear_1981}%
  \BibitemOpen
  \bibfield  {author} {\bibinfo {author} {\bibfnamefont {G.~N.}\ \bibnamefont
  {Bochkov}}\ and\ \bibinfo {author} {\bibfnamefont {Y.~E.}\ \bibnamefont
  {Kuzovlev}},\ }\href {https://doi.org/10.1016/0378-4371(81)90122-9}
  {\bibfield  {journal} {\bibinfo  {journal} {Phys. A: Stat. Mech. Appl.}\
  }\textbf {\bibinfo {volume} {106}},\ \bibinfo {pages} {443} (\bibinfo {year}
  {1981})}\BibitemShut {NoStop}%
\bibitem [{\citenamefont {Stratonovich}(1992)}]{stratonovich_nonlinear_1992}%
  \BibitemOpen
  \bibfield  {author} {\bibinfo {author} {\bibfnamefont {R.~L.}\ \bibnamefont
  {Stratonovich}},\ }\href {https://doi.org/10.1007/978-3-642-77343-3} {\emph
  {\bibinfo {title} {Nonlinear {Nonequilibrium} {Thermodynamics} {I}: {Linear}
  and nonlinear fluctuation-dissipation theorems}}}\ (\bibinfo  {publisher}
  {Springer},\ \bibinfo {year} {1992})\BibitemShut {NoStop}%
\bibitem [{\citenamefont {Evans}\ and\ \citenamefont
  {Morriss}(2008)}]{evans_statistical_2008}%
  \BibitemOpen
  \bibfield  {author} {\bibinfo {author} {\bibfnamefont {D.~J.}\ \bibnamefont
  {Evans}}\ and\ \bibinfo {author} {\bibfnamefont {G.}~\bibnamefont
  {Morriss}},\ }\href {https://doi.org/10.1017/CBO9780511535307} {\emph
  {\bibinfo {title} {Statistical {Mechanics} of {Nonequilibrium} {Liquids}}}},\
  \bibinfo {edition} {2nd}\ ed.\ (\bibinfo  {publisher} {Cambridge University
  Press},\ \bibinfo {year} {2008})\BibitemShut {NoStop}%
\bibitem [{\citenamefont {Fuchs}\ and\ \citenamefont
  {Cates}(2005)}]{fuchs_integration_2005}%
  \BibitemOpen
  \bibfield  {author} {\bibinfo {author} {\bibfnamefont {M.}~\bibnamefont
  {Fuchs}}\ and\ \bibinfo {author} {\bibfnamefont {M.~E.}\ \bibnamefont
  {Cates}},\ }\href {https://doi.org/10.1088/0953-8984/17/20/003} {\bibfield
  {journal} {\bibinfo  {journal} {J. Phys.: Condens. Matter}\ }\textbf
  {\bibinfo {volume} {17}},\ \bibinfo {pages} {S1681} (\bibinfo {year}
  {2005})}\BibitemShut {NoStop}%
\bibitem [{\citenamefont {Holsten}\ and\ \citenamefont
  {Krüger}(2021)}]{holsten_thermodynamic_2021}%
  \BibitemOpen
  \bibfield  {author} {\bibinfo {author} {\bibfnamefont {T.}~\bibnamefont
  {Holsten}}\ and\ \bibinfo {author} {\bibfnamefont {M.}~\bibnamefont
  {Krüger}},\ }\href {https://doi.org/10.1103/PhysRevE.103.032116} {\bibfield
  {journal} {\bibinfo  {journal} {Phys. Rev. E}\ }\textbf {\bibinfo {volume}
  {103}},\ \bibinfo {pages} {032116} (\bibinfo {year} {2021})}\BibitemShut
  {NoStop}%
\bibitem [{\citenamefont {Oppenheim}(1989)}]{oppenheim_nonlinear_1989}%
  \BibitemOpen
  \bibfield  {author} {\bibinfo {author} {\bibfnamefont {I.}~\bibnamefont
  {Oppenheim}},\ }\href {https://doi.org/10.1143/PTPS.99.369} {\bibfield
  {journal} {\bibinfo  {journal} {Prog. Theor. Phys. Supp.}\ }\textbf {\bibinfo
  {volume} {99}},\ \bibinfo {pages} {369} (\bibinfo {year} {1989})}\BibitemShut
  {NoStop}%
\bibitem [{\citenamefont {Bouchaud}\ and\ \citenamefont
  {Biroli}(2005)}]{bouchaud_nonlinear_2005}%
  \BibitemOpen
  \bibfield  {author} {\bibinfo {author} {\bibfnamefont {J.-P.}\ \bibnamefont
  {Bouchaud}}\ and\ \bibinfo {author} {\bibfnamefont {G.}~\bibnamefont
  {Biroli}},\ }\href {https://doi.org/10.1103/PhysRevB.72.064204} {\bibfield
  {journal} {\bibinfo  {journal} {Phys. Rev. B}\ }\textbf {\bibinfo {volume}
  {72}},\ \bibinfo {pages} {064204} (\bibinfo {year} {2005})}\BibitemShut
  {NoStop}%
\bibitem [{\citenamefont {Lippiello}\ \emph {et~al.}(2008)\citenamefont
  {Lippiello}, \citenamefont {Corberi}, \citenamefont {Sarracino},\ and\
  \citenamefont {Zannetti}}]{lippiello_nonlinear_2008}%
  \BibitemOpen
  \bibfield  {author} {\bibinfo {author} {\bibfnamefont {E.}~\bibnamefont
  {Lippiello}}, \bibinfo {author} {\bibfnamefont {F.}~\bibnamefont {Corberi}},
  \bibinfo {author} {\bibfnamefont {A.}~\bibnamefont {Sarracino}},\ and\
  \bibinfo {author} {\bibfnamefont {M.}~\bibnamefont {Zannetti}},\ }\href
  {https://doi.org/10.1103/PhysRevE.78.041120} {\bibfield  {journal} {\bibinfo
  {journal} {Phys. Rev. E}\ }\textbf {\bibinfo {volume} {78}},\ \bibinfo
  {pages} {041120} (\bibinfo {year} {2008})}\BibitemShut {NoStop}%
\bibitem [{\citenamefont {Andrieux}\ and\ \citenamefont
  {Gaspard}(2008)}]{andrieux_quantum_2008}%
  \BibitemOpen
  \bibfield  {author} {\bibinfo {author} {\bibfnamefont {D.}~\bibnamefont
  {Andrieux}}\ and\ \bibinfo {author} {\bibfnamefont {P.}~\bibnamefont
  {Gaspard}},\ }\href {https://doi.org/10.1103/PhysRevLett.100.230404}
  {\bibfield  {journal} {\bibinfo  {journal} {Phys. Rev. Lett.}\ }\textbf
  {\bibinfo {volume} {100}},\ \bibinfo {pages} {230404} (\bibinfo {year}
  {2008})}\BibitemShut {NoStop}%
\bibitem [{\citenamefont {Lucarini}\ and\ \citenamefont
  {Colangeli}(2012)}]{lucarini_beyond_2012}%
  \BibitemOpen
  \bibfield  {author} {\bibinfo {author} {\bibfnamefont {V.}~\bibnamefont
  {Lucarini}}\ and\ \bibinfo {author} {\bibfnamefont {M.}~\bibnamefont
  {Colangeli}},\ }\href {https://doi.org/10.1088/1742-5468/2012/05/P05013}
  {\bibfield  {journal} {\bibinfo  {journal} {J. Stat. Mech.}\ }\textbf
  {\bibinfo {volume} {2012}},\ \bibinfo {pages} {P05013} (\bibinfo {year}
  {2012})}\BibitemShut {NoStop}%
\bibitem [{\citenamefont {Diezemann}(2012)}]{diezemann_nonlinear_2012}%
  \BibitemOpen
  \bibfield  {author} {\bibinfo {author} {\bibfnamefont {G.}~\bibnamefont
  {Diezemann}},\ }\href {https://doi.org/10.1103/PhysRevE.85.051502} {\bibfield
   {journal} {\bibinfo  {journal} {Phys. Rev. E}\ }\textbf {\bibinfo {volume}
  {85}},\ \bibinfo {pages} {051502} (\bibinfo {year} {2012})}\BibitemShut
  {NoStop}%
\bibitem [{\citenamefont {Wang}\ and\ \citenamefont
  {Heinz}(2002)}]{wang_generalized_2002}%
  \BibitemOpen
  \bibfield  {author} {\bibinfo {author} {\bibfnamefont {E.}~\bibnamefont
  {Wang}}\ and\ \bibinfo {author} {\bibfnamefont {U.}~\bibnamefont {Heinz}},\
  }\href {https://doi.org/10.1103/PhysRevD.66.025008} {\bibfield  {journal}
  {\bibinfo  {journal} {Phys. Rev. D}\ }\textbf {\bibinfo {volume} {66}},\
  \bibinfo {pages} {025008} (\bibinfo {year} {2002})}\BibitemShut {NoStop}%
\bibitem [{\citenamefont {Andrieux}\ and\ \citenamefont
  {Gaspard}(2007)}]{andrieux_fluctuation_2007}%
  \BibitemOpen
  \bibfield  {author} {\bibinfo {author} {\bibfnamefont {D.}~\bibnamefont
  {Andrieux}}\ and\ \bibinfo {author} {\bibfnamefont {P.}~\bibnamefont
  {Gaspard}},\ }\href {https://doi.org/10.1088/1742-5468/2007/02/P02006}
  {\bibfield  {journal} {\bibinfo  {journal} {J. Stat. Mech.}\ }\textbf
  {\bibinfo {volume} {2007}},\ \bibinfo {pages} {P02006} (\bibinfo {year}
  {2007})}\BibitemShut {NoStop}%
\bibitem [{\citenamefont {Colangeli}\ \emph {et~al.}(2011)\citenamefont
  {Colangeli}, \citenamefont {Maes},\ and\ \citenamefont
  {Wynants}}]{colangeli_meaningful_2011}%
  \BibitemOpen
  \bibfield  {author} {\bibinfo {author} {\bibfnamefont {M.}~\bibnamefont
  {Colangeli}}, \bibinfo {author} {\bibfnamefont {C.}~\bibnamefont {Maes}},\
  and\ \bibinfo {author} {\bibfnamefont {B.}~\bibnamefont {Wynants}},\ }\href
  {https://doi.org/10.1088/1751-8113/44/9/095001} {\bibfield  {journal}
  {\bibinfo  {journal} {J. Phys. A: Math. Theor.}\ }\textbf {\bibinfo {volume}
  {44}},\ \bibinfo {pages} {095001} (\bibinfo {year} {2011})}\BibitemShut
  {NoStop}%
\bibitem [{\citenamefont {Basu}\ \emph {et~al.}(2015)\citenamefont {Basu},
  \citenamefont {Krüger}, \citenamefont {Lazarescu},\ and\ \citenamefont
  {Maes}}]{basu_frenetic_2015}%
  \BibitemOpen
  \bibfield  {author} {\bibinfo {author} {\bibfnamefont {U.}~\bibnamefont
  {Basu}}, \bibinfo {author} {\bibfnamefont {M.}~\bibnamefont {Krüger}},
  \bibinfo {author} {\bibfnamefont {A.}~\bibnamefont {Lazarescu}},\ and\
  \bibinfo {author} {\bibfnamefont {C.}~\bibnamefont {Maes}},\ }\href
  {https://doi.org/10.1039/C4CP04977B} {\bibfield  {journal} {\bibinfo
  {journal} {Phys. Chem. Chem. Phys.}\ }\textbf {\bibinfo {volume} {17}},\
  \bibinfo {pages} {6653} (\bibinfo {year} {2015})}\BibitemShut {NoStop}%
\bibitem [{\citenamefont {Basu}\ \emph {et~al.}(2018)\citenamefont {Basu},
  \citenamefont {Helden},\ and\ \citenamefont
  {Kr\"uger}}]{basu_extrapolation_2018}%
  \BibitemOpen
  \bibfield  {author} {\bibinfo {author} {\bibfnamefont {U.}~\bibnamefont
  {Basu}}, \bibinfo {author} {\bibfnamefont {L.}~\bibnamefont {Helden}},\ and\
  \bibinfo {author} {\bibfnamefont {M.}~\bibnamefont {Kr\"uger}},\ }\href
  {https://doi.org/10.1103/PhysRevLett.120.180604} {\bibfield  {journal}
  {\bibinfo  {journal} {Phys. Rev. Lett.}\ }\textbf {\bibinfo {volume} {120}},\
  \bibinfo {pages} {180604} (\bibinfo {year} {2018})}\BibitemShut {NoStop}%
\bibitem [{\citenamefont {Maes}(2020{\natexlab{a}})}]{maes_response_2020}%
  \BibitemOpen
  \bibfield  {author} {\bibinfo {author} {\bibfnamefont {C.}~\bibnamefont
  {Maes}},\ }\href
  {https://www.frontiersin.org/articles/10.3389/fphy.2020.00229} {\bibfield
  {journal} {\bibinfo  {journal} {Front. Phys.}\ }\textbf {\bibinfo {volume}
  {8}} (\bibinfo {year} {2020}{\natexlab{a}})}\BibitemShut {NoStop}%
\bibitem [{\citenamefont {Caspers}\ and\ \citenamefont
  {Krüger}(2024)}]{caspers_nonlinear_2024}%
  \BibitemOpen
  \bibfield  {author} {\bibinfo {author} {\bibfnamefont {J.}~\bibnamefont
  {Caspers}}\ and\ \bibinfo {author} {\bibfnamefont {M.}~\bibnamefont
  {Krüger}},\ }\href {https://doi.org/10.1063/5.0227674} {\bibfield  {journal}
  {\bibinfo  {journal} {J. Chem. Phys.}\ }\textbf {\bibinfo {volume} {161}},\
  \bibinfo {pages} {124109} (\bibinfo {year} {2024})}\BibitemShut {NoStop}%
\bibitem [{\citenamefont {Helden}\ \emph {et~al.}(2016)\citenamefont {Helden},
  \citenamefont {Basu}, \citenamefont {Krüger},\ and\ \citenamefont
  {Bechinger}}]{helden_measurement_2016}%
  \BibitemOpen
  \bibfield  {author} {\bibinfo {author} {\bibfnamefont {L.}~\bibnamefont
  {Helden}}, \bibinfo {author} {\bibfnamefont {U.}~\bibnamefont {Basu}},
  \bibinfo {author} {\bibfnamefont {M.}~\bibnamefont {Krüger}},\ and\ \bibinfo
  {author} {\bibfnamefont {C.}~\bibnamefont {Bechinger}},\ }\href
  {https://doi.org/10.1209/0295-5075/116/60003} {\bibfield  {journal} {\bibinfo
   {journal} {EPL}\ }\textbf {\bibinfo {volume} {116}},\ \bibinfo {pages}
  {60003} (\bibinfo {year} {2016})}\BibitemShut {NoStop}%
\bibitem [{\citenamefont {Maes}(2014)}]{maes_second_2014}%
  \BibitemOpen
  \bibfield  {author} {\bibinfo {author} {\bibfnamefont {C.}~\bibnamefont
  {Maes}},\ }\href {https://doi.org/10.1007/s10955-013-0904-8} {\bibfield
  {journal} {\bibinfo  {journal} {J. Stat. Phys.}\ }\textbf {\bibinfo {volume}
  {154}},\ \bibinfo {pages} {705} (\bibinfo {year} {2014})}\BibitemShut
  {NoStop}%
\bibitem [{\citenamefont {Maes}(2020{\natexlab{b}})}]{maes_frenesy_2020}%
  \BibitemOpen
  \bibfield  {author} {\bibinfo {author} {\bibfnamefont {C.}~\bibnamefont
  {Maes}},\ }\href {https://doi.org/10.1016/j.physrep.2020.01.002} {\bibfield
  {journal} {\bibinfo  {journal} {Phys. Rep.}\ }\textbf {\bibinfo {volume}
  {850}},\ \bibinfo {pages} {1} (\bibinfo {year}
  {2020}{\natexlab{b}})}\BibitemShut {NoStop}%
\bibitem [{\citenamefont {Gomez-Solano}\ \emph {et~al.}(2009)\citenamefont
  {Gomez-Solano}, \citenamefont {Petrosyan}, \citenamefont {Ciliberto},
  \citenamefont {Chetrite},\ and\ \citenamefont
  {Gawedzki}}]{gomez-solano_experimental_2009}%
  \BibitemOpen
  \bibfield  {author} {\bibinfo {author} {\bibfnamefont {J.~R.}\ \bibnamefont
  {Gomez-Solano}}, \bibinfo {author} {\bibfnamefont {A.}~\bibnamefont
  {Petrosyan}}, \bibinfo {author} {\bibfnamefont {S.}~\bibnamefont
  {Ciliberto}}, \bibinfo {author} {\bibfnamefont {R.}~\bibnamefont
  {Chetrite}},\ and\ \bibinfo {author} {\bibfnamefont {K.}~\bibnamefont
  {Gawedzki}},\ }\href {https://doi.org/10.1103/PhysRevLett.103.040601}
  {\bibfield  {journal} {\bibinfo  {journal} {Phys. Rev. Lett.}\ }\textbf
  {\bibinfo {volume} {103}},\ \bibinfo {pages} {040601} (\bibinfo {year}
  {2009})}\BibitemShut {NoStop}%
\bibitem [{\citenamefont {Blickle}\ \emph {et~al.}(2007)\citenamefont
  {Blickle}, \citenamefont {Speck}, \citenamefont {Lutz}, \citenamefont
  {Seifert},\ and\ \citenamefont {Bechinger}}]{blickle_einstein_2007}%
  \BibitemOpen
  \bibfield  {author} {\bibinfo {author} {\bibfnamefont {V.}~\bibnamefont
  {Blickle}}, \bibinfo {author} {\bibfnamefont {T.}~\bibnamefont {Speck}},
  \bibinfo {author} {\bibfnamefont {C.}~\bibnamefont {Lutz}}, \bibinfo {author}
  {\bibfnamefont {U.}~\bibnamefont {Seifert}},\ and\ \bibinfo {author}
  {\bibfnamefont {C.}~\bibnamefont {Bechinger}},\ }\href
  {https://doi.org/10.1103/PhysRevLett.98.210601} {\bibfield  {journal}
  {\bibinfo  {journal} {Phys. Rev. Lett.}\ }\textbf {\bibinfo {volume} {98}},\
  \bibinfo {pages} {210601} (\bibinfo {year} {2007})}\BibitemShut {NoStop}%
\bibitem [{\citenamefont {Mehl}\ \emph {et~al.}(2010)\citenamefont {Mehl},
  \citenamefont {Blickle}, \citenamefont {Seifert},\ and\ \citenamefont
  {Bechinger}}]{mehl_experimental_2010}%
  \BibitemOpen
  \bibfield  {author} {\bibinfo {author} {\bibfnamefont {J.}~\bibnamefont
  {Mehl}}, \bibinfo {author} {\bibfnamefont {V.}~\bibnamefont {Blickle}},
  \bibinfo {author} {\bibfnamefont {U.}~\bibnamefont {Seifert}},\ and\ \bibinfo
  {author} {\bibfnamefont {C.}~\bibnamefont {Bechinger}},\ }\href
  {https://doi.org/10.1103/PhysRevE.82.032401} {\bibfield  {journal} {\bibinfo
  {journal} {Phys. Rev. E}\ }\textbf {\bibinfo {volume} {82}},\ \bibinfo
  {pages} {032401} (\bibinfo {year} {2010})}\BibitemShut {NoStop}%
\bibitem [{\citenamefont {Gomez-Solano}\ \emph {et~al.}(2011)\citenamefont
  {Gomez-Solano}, \citenamefont {Petrosyan}, \citenamefont {Ciliberto},\ and\
  \citenamefont {Maes}}]{gomez-solano_fluctuations_2011}%
  \BibitemOpen
  \bibfield  {author} {\bibinfo {author} {\bibfnamefont {J.~R.}\ \bibnamefont
  {Gomez-Solano}}, \bibinfo {author} {\bibfnamefont {A.}~\bibnamefont
  {Petrosyan}}, \bibinfo {author} {\bibfnamefont {S.}~\bibnamefont
  {Ciliberto}},\ and\ \bibinfo {author} {\bibfnamefont {C.}~\bibnamefont
  {Maes}},\ }\href {https://doi.org/10.1088/1742-5468/2011/01/P01008}
  {\bibfield  {journal} {\bibinfo  {journal} {J. Stat. Mech.}\ }\textbf
  {\bibinfo {volume} {2011}},\ \bibinfo {pages} {P01008} (\bibinfo {year}
  {2011})}\BibitemShut {NoStop}%
\bibitem [{\citenamefont {Bochkov}\ and\ \citenamefont
  {Kuzovlev}(1977)}]{bochkov_general_1977}%
  \BibitemOpen
  \bibfield  {author} {\bibinfo {author} {\bibfnamefont {G.~N.}\ \bibnamefont
  {Bochkov}}\ and\ \bibinfo {author} {\bibfnamefont {Y.~E.}\ \bibnamefont
  {Kuzovlev}},\ }\href {http://www.jetp.ras.ru/cgi-bin/dn/e_045_01_0125.pdf}
  {\bibfield  {journal} {\bibinfo  {journal} {Zh. Eksp. Teor. Fiz.}\ }\textbf
  {\bibinfo {volume} {72}},\ \bibinfo {pages} {238} (\bibinfo {year}
  {1977})}\BibitemShut {NoStop}%
\bibitem [{\citenamefont {Jarzynski}(1997)}]{jarzynski_nonequilibrium_1997}%
  \BibitemOpen
  \bibfield  {author} {\bibinfo {author} {\bibfnamefont {C.}~\bibnamefont
  {Jarzynski}},\ }\href {https://doi.org/10.1103/PhysRevLett.78.2690}
  {\bibfield  {journal} {\bibinfo  {journal} {Phys. Rev. Lett.}\ }\textbf
  {\bibinfo {volume} {78}},\ \bibinfo {pages} {2690} (\bibinfo {year}
  {1997})}\BibitemShut {NoStop}%
\bibitem [{\citenamefont {Crooks}(1999)}]{crooks_entropy_1999}%
  \BibitemOpen
  \bibfield  {author} {\bibinfo {author} {\bibfnamefont {G.~E.}\ \bibnamefont
  {Crooks}},\ }\href {https://doi.org/10.1103/PhysRevE.60.2721} {\bibfield
  {journal} {\bibinfo  {journal} {Phys. Rev. E}\ }\textbf {\bibinfo {volume}
  {60}},\ \bibinfo {pages} {2721} (\bibinfo {year} {1999})}\BibitemShut
  {NoStop}%
\bibitem [{\citenamefont {Seifert}(2012)}]{seifert_stochastic_2012}%
  \BibitemOpen
  \bibfield  {author} {\bibinfo {author} {\bibfnamefont {U.}~\bibnamefont
  {Seifert}},\ }\href {https://doi.org/10.1088/0034-4885/75/12/126001}
  {\bibfield  {journal} {\bibinfo  {journal} {Rep. Prog. Phys.}\ }\textbf
  {\bibinfo {volume} {75}},\ \bibinfo {pages} {126001} (\bibinfo {year}
  {2012})}\BibitemShut {NoStop}%
\bibitem [{\citenamefont {Caspers}\ and\ \citenamefont
  {Kr\"uger}(2025)}]{caspers_companion}%
  \BibitemOpen
  \bibfield  {author} {\bibinfo {author} {\bibfnamefont {J.}~\bibnamefont
  {Caspers}}\ and\ \bibinfo {author} {\bibfnamefont {M.}~\bibnamefont
  {Kr\"uger}},\ }\href {https://doi.org/10.1103/pzs6-l3ws} {\bibfield
  {journal} {\bibinfo  {journal} {Phys. Rev. E}\ }\textbf {\bibinfo {volume}
  {112}},\ \bibinfo {pages} {024124} (\bibinfo {year} {2025})}\BibitemShut
  {NoStop}%
\bibitem [{\citenamefont {Maes}(2021)}]{maes_local_2021}%
  \BibitemOpen
  \bibfield  {author} {\bibinfo {author} {\bibfnamefont {C.}~\bibnamefont
  {Maes}},\ }\href {https://doi.org/10.21468/SciPostPhysLectNotes.32}
  {\bibfield  {journal} {\bibinfo  {journal} {SciPost Phys. Lect. Notes}\ ,\
  \bibinfo {pages} {32}} (\bibinfo {year} {2021})}\BibitemShut {NoStop}%
\bibitem [{Note4()}]{Note4}%
  \BibitemOpen
  \bibinfo {note} {The expansion in Eq.~\protect \eqref {eq:NoiseIdentity}
  suggests the dimensionless expansion parameter $\beta F \DOTSI \intop
  \ilimits@ _{t-\tau }^t \protect \mathrm {d}s\protect \, \protect \dot x_s$
  with cumulant relaxation time $\tau $}\BibitemShut {NoStop}%
\bibitem [{\citenamefont {Cates}\ and\ \citenamefont
  {Candau}(1990)}]{cates_statics_1990}%
  \BibitemOpen
  \bibfield  {author} {\bibinfo {author} {\bibfnamefont {M.~E.}\ \bibnamefont
  {Cates}}\ and\ \bibinfo {author} {\bibfnamefont {S.~J.}\ \bibnamefont
  {Candau}},\ }\href {https://doi.org/10.1088/0953-8984/2/33/001} {\bibfield
  {journal} {\bibinfo  {journal} {J. Phys.: Condens. Matter}\ }\textbf
  {\bibinfo {volume} {2}},\ \bibinfo {pages} {6869} (\bibinfo {year}
  {1990})}\BibitemShut {NoStop}%
\bibitem [{\citenamefont {Gomez-Solano}\ and\ \citenamefont
  {Bechinger}(2015)}]{gomez-solano_transient_2015}%
  \BibitemOpen
  \bibfield  {author} {\bibinfo {author} {\bibfnamefont {J.~R.}\ \bibnamefont
  {Gomez-Solano}}\ and\ \bibinfo {author} {\bibfnamefont {C.}~\bibnamefont
  {Bechinger}},\ }\href {https://doi.org/10.1088/1367-2630/17/10/103032}
  {\bibfield  {journal} {\bibinfo  {journal} {New J. Phys.}\ }\textbf {\bibinfo
  {volume} {17}},\ \bibinfo {pages} {103032} (\bibinfo {year}
  {2015})}\BibitemShut {NoStop}%
\bibitem [{Note2()}]{Note2}%
  \BibitemOpen
  \bibinfo {note} {$\omega $ is determined from the power spectral density,
  thus carrying an error depending on the length of the
  measurement.}\BibitemShut {Stop}%
\bibitem [{Note3()}]{Note3}%
  \BibitemOpen
  \bibinfo {note} {As the phases $\phi _2^{(1)}$ and $\phi _2^{(2)}$ may
  differ, the coefficient $\Delta A_2$ of the difference of first and second
  cumulants is found via $\Delta A_2 \equiv \protect \sqrt {\left (A_2^{(1)}
  \right )^2+\left (A_2^{(2)}\right )^2-2 A_2^{(1)} A_2^{(2)}\cos \left (\phi
  _2^{(1)}-\phi _2^{(2)}\right )}$. For $m=0$, i.e., the zero frequency
  contribution, there is no phase by definition, and the coefficients
  $A_0^{(n)}$ can be compared directly}\BibitemShut {NoStop}%
\end{thebibliography}
%

\end{document}